# Time vs. Ensemble Averages for Nonstationary Time Series

Joseph L. McCauley
Physics Department
University of Houston
Houston, Tx. 77204
jmccauley@uh.edu

## Abstract

We analyze the question whether sliding window time averages applied to stationary increment processes converge to a limit (in probability). The question centers on averages, correlations, and densities constructed via time averages of the increment $x(t,T)=x(t+T)-x(t)$, e.g. $x(t,T)=\ln(p(t+T)/p(t))$ in finance and economics where $p(t)$ is a price, and the assumption is that the increment is distributed independently of t. We show that the condition for applying Thebyshev's Theorem to time averages of functions of stationary increments is strongly violated. We argue that, for both stationary and nonstationary increments, Tchebyshev's Theorem provides the basis for constructing ensemble averages and densities from a single, historic time series if, as in FX markets, the series shows a definite 'statistical periodicity' on the average.

# 1. Introduction

Continuous stationary processes x(t) have the nice property that the limit of the time average of x(t) exists in probability, a generalization of the law of large numbers [1]. That limit is generally unknown. If, in addition, the pair correlations have the right asymptotic behavior, then the process is ergodic: time averages converge with probability one to ensemble averages [1]. A weakly stationary process is one where the 1-point and 2-point probability densities are time translationally invariant, so that the mean and variance are constants, and the pair correlations obey $R(T) = \langle x(t)x(t+T) \rangle = \langle x(0)x(T) \rangle$. With the average subtracted from x(t), the condition in discrete time for ergodicity is that $R(T) \to 0$ as $T \to \infty$ [1]. For continuous time processes one needs $T^{-1} \int_0^T R(s)ds \to 0, T \to \infty$ [1,2,3].

Historically, ergodicity in deterministic dynamical systems was important for statistical physics in order to justify the equal a priori probabilities assigned in phase space by Boltzmann and Gibbs [4,5]. Ergodicity would guarantee that a certain statistical ensemble, Gibbs' microcanonical ensemble, could be relied on to produce correct results.

In economics and finance, stationarity of time series has been desired because there is no obvious statistical ensemble, instead of the reruns of identical experiments that define a statistical ensemble, one is faced with a single, historic time series. A financial price series for a stock provides an example. An ensemble, in contrast, provides a collection of statistically identical time series from which histograms and averages can be determined for each time t. It was argued by Bassler et al [6] that ensembles are necessary for finance analysis and were constructed. Examples were provided where both the process and the increments are nonstationary.

Financial time series are strongly nonstationary [7,8,9]. The 'solution' to the problem of data analysis, in lieu of constructing ensembles, has been to slide a window along a single, historic time series to construct averages and histograms, using the increment $x(t,T)=\ln(p(t+T)/p(t))$ as variable [7,8,9,10]. These constructions are inherently time averages, but the question whether the time averages so constructed converge in principle to a limit (in probability) has not been addressed. Stationarity of increments of financial time series is widely and implicitly assumed by equating scaling to an assumption of long time correlations [11], but no test for increment stationarity was performed to justify that assumption in any paper in the literature, so far as we can see.

In this paper we do not discuss further whether increments in financial time series are or are not stationary [6,12]. Here, we focus soley on the question: given a nonstationary process with stationary increments, does the sliding window method generate averages and histograms that can be guaranteed to converge to a limit 'in probability'? In particular, can we guarantee convergence to the respective ensemble limits. The use of the sliding window is equivalent to treating the increment $x(t,T)=\ln(p(t+T)/p(t))$ as a variable in a stochastic process [12]. This is taken for granted in econophysics in particular, and in time series analysis in general [10].

Our stated task requires a convergence theorem that can be applied to time averages based on stationary increments. There are exactly three classes of convergence theorem available to us in probability theory: Tchebyshev's Theorem (the law of large numbers), the Central Limit Theorem, and martingale convergence theorems. The second is more restrictive than the first, and the first is adequate for the task.

## 2. Stationary Increment Processes

Consider a nonstationary process x(t) with stationary increments. Stationary increments [2,12,14,15] means that the increment relation

$$x(t,T) = x(t+T) - x(t) = x(0,T) \qquad (1)$$

holds 'in probability'. Let $f_n(x_n,t_n;\ldots;x_1,t_1)$ denote the n-point density of the process, n=1,2,3,… . Equality 'in probability', with z=x(t+T)-x(t)=y-x, means in this case that the 1-point increment density

$$f(z,t,t+T) = \iint dy dx f_2(y,t+T;x,t)\delta(z-y+x) \qquad (2)$$

or

$$f(z,t,t+T) = \int dx f_2(x+z,t+T;x,t) \qquad (3)$$

is independent of t, depends on the lag time T alone, f(z,t,t+T)=f(z,0,T). There is absolutely no requirement placed on time translational invariance of the densities $f_n$, n>2, so that the pair correlations of stationary increment processes are not restricted. E.g., fBm [14,15] has stationary increments and long range increment autocorrelations $\langle x(t,T)x(t,-T)\rangle \neq 0$, whereas the Wiener process has stationary increments and vanishing increment autocorrelations $\langle x(t,T)x(t,-T)\rangle = 0$ [12,15].

For drift free diffusive processes (ruling out fBm), processes with vanishing increment autocorrelations for nonoverlapping time intervals, a stationary increment process is identical with a time-translationally invariant martingale, a process where the diffusion coefficient D(x,t) depends on x alone [16]. Such processes are generally nonstationary [17]. In 'weak increment stationarity' of a diffusive process one requires only that the mean square fluctuation is time translationally invariant, that $\langle x^2(t,T) \rangle = \langle x^2(0,T) \rangle = \langle x^2(1) \rangle T$. For diffusive processes (Ito processes) this admits nonstationary increment processes like a martingale scaling with a Hurst exponent H=1/2 [18].

Stationarity of increments can be tested by the construction of ensemble averages from a single, historic time series. In fact, there is no other known way to establish that increments are stationary. Our question is, however: if stationarity of increments were established via ensemble averages of data, or the choice a time translationally invariant Ito model (examples are diffusive models with $D(x) = x, D(x) = 1+|x|$), can one justify using the sliding window technique to generate results as time averages? Stated otherwise, to what limits should the sliding window averages and histograms converge? There is possible ambiguity because with stationary increments the increment density is exactly (3) but this is generally not the same as the 1-point density

$$f_1(x,T) = \int dy f_2(x,T;y,0), \qquad (4)$$

even though z=x(t,T)=x(0,T)=x 'in probability' if we take x(0)=0. So to which limit does should a sliding window histogram converge? We will now show that the assumption of a definite limit is not justified by the existing theory.

## 3. Sliding Window Time Averages

We begin on a positive note. If $x(t,T)=x(t+T)-x(t)=x(0,T)$ 'in probability', and if the increments are uncorrelated (ruling out fBm) then by Tchebyshev's Theorem [1] time averages of $x(t,T)$

$$\langle x(t,T) \rangle_{timeavg} = \frac{1}{N} \sum_{k=1}^{N} x(kT,T) \qquad (5)$$

converge in probability to the ensemble average value $\langle x(0,T) \rangle = 0$, if as we have done in (5), we restrict to times $t=nT$, n integral, so that the random variables in the summand have no pair correlations.

However, Tchebyshev's Theorem cannot be applied to predict that time averages of the mean square fluctuation,

$$\langle x^2(t,T) \rangle_{timeavg} = \frac{1}{N} \sum_{k=1}^{N} x^2(kT,T), \qquad (6)$$

converge to the ensemble average result $\langle x^2(0,T) \rangle = \langle x^2(1) \rangle T$, because mean square fluctuations for nonoverlapping time intervals are not necessarily uncorrelated ($\langle x^2(t,T)x^2(t,-T) \rangle$ is a 'volatility measure'). Eqn. (6) is exactly the formula for a sliding window calculation of the mean square fluctuation [10].

What about the histograms obtained from sliding a window on a stationary increment time series? Do the histograms converge to any particular limit in probability?

The time average of the increment density is defined as

$$f_s(z,T) = \frac{1}{N} \sum_{t=t_1}^{t_N} \delta(z - x(t,T)) \qquad (7)$$

where in this case the delta function is the Kronecker delta. The subscript s means "sliding window" average. Since N is the number of points in the time series, to insure uncorrelated increments in (5) we should restrict to t=nT, but this doesn't matter. Note that the definition of the 2-point ensemble average increment density is

$$f(z_1,t_1,t_1+T;z_2,t_2,t_2+T) = \langle \delta(z_1 - x(t_1,T))\delta(z_2 - x(t_2,T)) \rangle. \qquad (8)$$

First, (i) the density defined by (8) is not necessarily time-translationally invariant, and (ii) this density generally doesn't vanish. The deltas in the time average (7) are therefore correlated, so that Tchebyshev's theorem does not apply. This means that we don't know if the time series (7) has any limit at all as N goes to infinity, much less a limit given by f or $f_1$. Hence, when histograms are constructed by sliding a window, there is no reason to expect that one has obtained either f(z,0,T) or $f_1$(z,T).

The ensemble average (8) yields

$$f(x,t,t+T;z',t',t'+T) = \int \prod_{k=1}^{4} dx_k \delta(z - x_4 + x_3)\delta(z' - x_2 + x_1) f_4(x_4,t+T;...,x_1,t')$$
(9)

which reduces to

$$f(x,t,t+T;z',t',t'+T) = \int dx_3 d x_1 f_2(x_2 + z, t+T, x_3, t; x_1 + z', t'+T, x_1, t') . \qquad (10)$$

For stationary increments this density neither vanishes nor, in general, factors into a product of independent 1-point densities. So stationary increments do not permit us to escape the necessity to construct and compute averages and densities from a statistical ensemble in the analysis of a single, historic time series. If a single historic time series cannot be broken up sensibly into N subseries to construct a statistical ensemble (the analog of rerunning the same experiment N times), then a data analysis may yield spurious results. The standard estimation procedures of statistics are not an alternative to statistical ensembles, and we've discussed elsewhere [16] the weakness in the methods of econometrics.

If we would try to appeal to Tchebyshev's Theorem applied to i.i.d. increments, then among processes with uncorrelated increments (martingales) only the Wiener process possesses the combined time and space translational invariances required to turn the Markov condition seen as products of n 1-point transition densities into an i.i.d. condition on the increments [16].

We've commented elsewhere [16] on the flaw in using the lack of convergence of a time average of mean square fluctuations of returns (6) to deduce fat tails in cotton price returns [10].

## 4. Ensemble Averages

Tchebyshev's Theorem provides the basis for ensemble averages based on N identical reruns of the same experiment. In what follows the process is generally nonstationary with nonstationary increments, there is no restriction to any sort of stationarity.

Let there be N experimental realizations of a time series x(t), where the system is strobbed at the same time sequence in each run. Consider the N points $x_k(t)$, k=1,…,N, for the runs

at the same time t. Then the histogram for the 1-point density is given by

$$f_1(x,t) \approx \frac{1}{N}\sum_{k=1}^{N}\delta(x-x_k) \qquad (11)$$

and will show scatter so long as N is finite, which is necessarily the case in experiment and simulations. To apply Tchebyshev's Theorem for convergence, we need

$$\langle \delta(x-x(t))\delta(y-x(t))\rangle = 0. \qquad (12)$$

The ensemble average is

$$\langle \delta(x-x(t))\delta(y-x(t))\rangle = \iint dx_1 dx_2 \delta(x-x_1)\delta(y-x_2)f_2(x_2,t;x_1,t).$$
(13)

With $f_2(y,t;x,s) = p_2(y,t|x,s)f_1(x,s)$ and $p_2(y,t|x,t) = \delta(y-x)$ we obtain (12) for y≠x, so the fixed time series for the histograms converges to the ensemble average $f_1(x,t)$. One can show similarly that the correlations of other quantities calculated at equal times vanishes similarly. Tchebyshev's Theorem provides the basis for the construction of statistical ensembles for general nonstationary processes.

In economics there is but a single historic time series. We've shown for finance data that a 'run' can be taken as one day of trading, in the sense that the statistics are periodic day after day to within scatter [6]. The weakness in the construction of that ensemble is that the starting prices each day are not strictly independent. Each trading day for FX markets runs 24 hours (5 days/week), and the clock was arbitrarily set at 9AM each day to start the new 'run'. So the first return of day n at time is the same as the last return of day n-1, and those two returns are martingale correlated,

$\langle x(t)x(t+T) \rangle = \langle x^2(t) \rangle$, and the relevant 'correlation' for the ensemble average (11) is in this case $f_2(x,t+T;y,t)$. With $f_2(x,t+T;y,t) = p_2(x,t+T|y,t)f_1(x,t)$ the question is whether $p_2(x,t+T|y,t) \ll 1, T \gg 1$? For a diffusive process we expect this to hold independently of x≠y, but making the conjecture rigorous is an unsolved problem. Perhaps the problem can be solved in the context of stopping times (e.g., the time average of the Wiener process transition density is the Coulomb potential and is connected to the fact that the stopping time for an unbounded interval is infinite [19]).

The periodicity on which finance market ensembles are based [6] was first noted in Gallucio, Caldarelli, Marsilli, and Zhang [20].

## Acknowledgement

JMC is grateful to Kevin Bassler, Gemunu Gunaratne, and Søren jJhansen for discussions.

## References

1. B.V. Gnedenko, *Theory of Probability*, transl. from Russian by B.D. Seckler, Chelsea, N.Y., 1968.

2. R.L. Stratonovich. *Topics in the Theory of Random Noise*, Gordon & Breach: N.Y., tr. R. A. Silverman, 1963.

3. A.M. Yaglom & I.M. Yaglom, *An introduction to the Theory of Stationary Random Functions*. Transl. and ed. by R. A. Silverman. Prentice-Hall, Englewood Cliffs, N.J., 1962.

4. M. Kac, *Statistical Independence on Probability, Number Theory and Analysis*, Carus Math. Monographs nr. 12, Wiley, Rahway, 1959.

5. P.R. Halmos, *Lectures on Ergodic Theory*, AMS Chelsea Publ., Providence, 1956.

6. K.E. Bassler, J. L. McCauley, & G.H. Gunaratne, *Nonstationary Increments, Scaling Distributions, and Variable Diffusion Processes in Financial Markets*, PNAS **104**, 172287, 2007.

7. R.E. Mantegna & H.E. Stanley, *Nature* **383**, 587, 1996.

8. M. Dacorogna et al., *An Intro. to High Frequency Finance*, Academic Pr., N.Y., 2001.

9. C. Renner, J. Peinke, & R. Friedrich, *Physica* **A298**, 49, 2001.

10. B. Mandelbrot, *J. Business* **36**, 420, 1963.

11. T. Di Matteo, T.Aste, & M.M. Dacorogna, *Physica* **A324**, 183, 2003.

12. K.E. Bassler, G.H. Gunaratne, and J.L. McCauley, *Empirically Based Modeling in Financial Economics and Beyond: Spurious Stylized Facts*, Int. Rev. Fin. An., 2008.

13. J.L. McCauley, K.E. Bassler, and G.H. Gunaratne, Martingales, Detranding Data, and the Efficient Market Hypothesis, Physica **A37**, 202, 2008.

14. Mandelbrot & J. W. van Ness, *SIAM Rev*. 10, 2, 422,1968.

15. P. Embrechts and M. Maejima, *Selfsimilar Processes*, Princeton University Press, Princeton, 2002.

16. J.L. McCauley, K.E. Bassler, & G.H. Gunaratne, *Integration I(d) of Nonstationary Time Series: Stationary and nonstationary increments*, preprint, 2008.

17. J.L. McCauley, *Nonstationarity of Efficient Financial Markets: FX Market Evolution from Stability to Instability*, Int. Rev. Fin. An., 2008.

18. K.E. Bassler, G.H. Gunaratne, & J. L. McCauley, *Hurst Exponents, Markov Processes, and Nonlinear Diffusion Equations*, Physica ***A369***, 343, 2006.

19. R. Durrett, *Brownian Motion and Martingales in Analysis*, Wadsworth, Belmont, 1984.

20. S. Gallucio, G. Caldarelli, M. Marsilli, and Y.-C. Zhang, *Physica **A245***, 423, 1997.